\documentclass[aps,prb,reprint,amsmath,amssymb,showpacs,superscriptaddress,twocolumn]{revtex4-1}
\usepackage{graphicx}
\usepackage{xcolor}
\usepackage{subfigure}
\usepackage{hyperref}
\usepackage{hypcap}
\usepackage{braket}
\usepackage{amsmath}
\usepackage{commath}
\usepackage{chemformula}
\usepackage{tabularx}

\newcommand{\bvec}[1]{\mathbf{#1}}

\newcommand{\xhat}{\bvec{\hat{x}}}

\newcommand{\mhat}{\bvec{\hat{m}}}

\newcommand{\TP}{(z,\phi)}
\newcommand{\TPp}{(z',\phi')}

\begin{document}
	\title{Unconventional spin-orbit torque in transition metal dichalcogenide/ferromagnet bilayers from first-principles calculations}
	\author{Fei Xue}	
	\affiliation{Physical Measurement Laboratory, National Institute of Standards and Technology, Gaithersburg, MD 20899, USA}
	\affiliation{Institute for Research in Electronics and Applied Physics \& Maryland Nanocenter,	University of Maryland, College Park, MD 20742}
	\author{Christoph Rohmann}
	\affiliation{Physical Measurement Laboratory, National Institute of Standards and Technology, Gaithersburg, MD 20899, USA}
	\affiliation{Institute for Research in Electronics and Applied Physics \& Maryland Nanocenter,	University of Maryland, College Park, MD 20742}
	\author{Junwen Li}
	\affiliation{DFTWorks LLC, Oakton, VA 22124}
	\author{Vivek Amin}
	\affiliation{Physical Measurement Laboratory, National Institute of Standards and Technology, Gaithersburg, MD 20899, USA}
	\affiliation{Institute for Research in Electronics and Applied Physics \& Maryland Nanocenter,	University of Maryland, College Park, MD 20742}
	\author{Paul Haney}	
	\affiliation{Physical Measurement Laboratory, National Institute of Standards and Technology, Gaithersburg, MD 20899, USA}
	\date{\today}
	
	\begin{abstract}
		Motivated by recent observations of unconventional out-of-plane dampinglike torque in \ch{WTe2}/Permalloy bilayer systems, we calculate the spin-orbit torque generated in two-dimensional transition metal dichalcogenide (TMD)-ferromagnet heterostructures using first-principles methods and linear response theory. Our numerical calculation of spin-orbit torques in \ch{WTe2}/Co and \ch{MoTe2}/Co heterostructures shows both conventional and novel dampinglike torkances (torque per electric field) with comparable magnitude, around $100 \hbar/2e~(\rm \Omega\cdot  cm)^{-1}$, for an electric field applied perpendicular to the mirror plane of the TMD layer. To gain further insight into the source of dampinglike torque, we compute the spin current flux between the TMD and Co layers and find good agreement between the two quantities.  This indicates that the conventional picture of dampinglike spin-orbit torque, whereby the torque results from the spin Hall effect plus spin transfer torque, largely applies to TMD/Co bilayer systems.
	\end{abstract}
	
	\maketitle
	\section{Introduction}
	
\begin{figure}[htbp]
		\includegraphics[width=0.9\columnwidth]{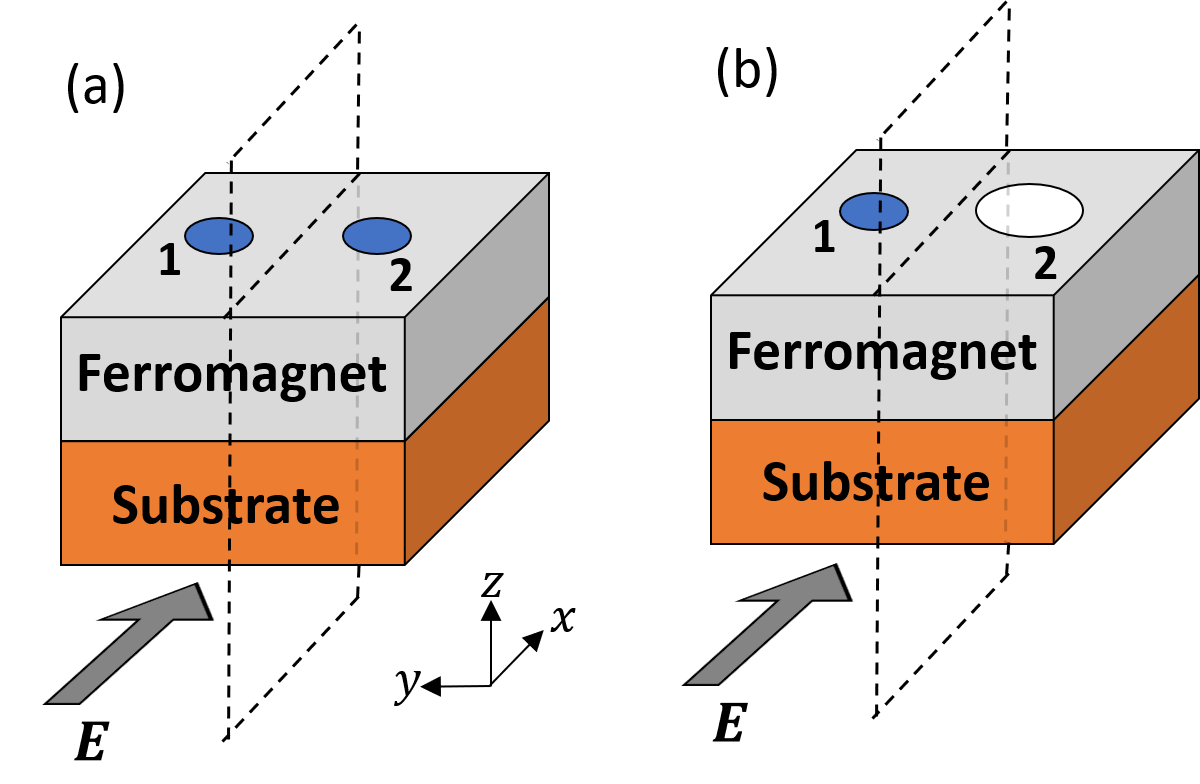}
		\caption{(a) Schematic of heavy metal (substrate)-ferromagnet bilayer with in-plane 4-fold symmetry; atoms 1 and 2 are equivalent and the system retains $y \rightarrow -y$ symmetry when the magnetization is along the ${\bf y}$ direction.  (b) Bilayer system with broken mirror symmetry; atoms 1 and 2 are inequivalent.  For magnetization along the ${\bf y}$ direction, there are no system symmetries and all components of torque are allowed.
		}
		\label{Fig:geom}
	\end{figure}

	Spin-orbit torque is an effect in which the application of an electric field induces the exchange of angular momentum between the crystal lattice and the magnetization of a magnetically ordered material \cite{manchon2009theory,liu2011spin,miron2011perpendicular,manchon2019current}. This exchange is mediated by spin-orbit coupling, and the effect offers a promising mechanism for energy-efficient electrical switching of thin film ferromagnets. The prototypical geometry of a spin-orbit torque-based device is shown in Fig.~\ref{Fig:geom}: charge current is applied in the plane of a heavy metal-ferromagnet bilayer, leading to magnetic dynamics and potentially switching of the ferromagnetic orientation.  This geometry is quite distinct from the conventional spin transfer torque-based magnetic tunnel junctions, which utilize spin-polarized current flowing perpendicular to the materials' interfaces \cite{ralph2008spin}.  The geometry of devices which utilize spin-orbit torque enables separate electrical paths for writing and reading the magnetic information, which may be advantageous for applications \cite{liu2012spin}.
	
	The dependence of the spin-orbit torque on the magnetization orientation is determined by the system symmetry.  In many devices studied to date ({\it e.g.}, Co-Pt bilayers), the materials are deposited by sputtering and the film is effectively isotropic in the plane normal to the interface.  This leaves the stacking direction (${\bf z}$) as the only broken-symmetry direction.  In the presence of an applied electric field ${\bf E}$, and for the magnetization aligned to the ${\bf E}\times {\bf z}$ direction, the system is invariant under reflections about the plane spanned by ${\bf E}$ and ${\bf z}$.  This implies that the torque on the magnetization vanishes in this configuration, so that it is a fixed point for electric field-induced magnetic dynamics \cite{Garello2013}.  This in turn implies that the spin-orbit torque for such a system can deterministically switch the magnetization between in-plane configurations \cite{liu2012spin,pai2012spin}.
	
	Breaking the in-plane symmetry of the system removes this constraint on the form of the torque.  In this case, the electric field-induced torque generally drives the magnetization to a point above or below the plane of the film (depending on the current polarity). This enables deterministic switching of perpendicularly magnetized thin film ferromagnets. This is of particular interest due to technological advantages of perpendicularly magnetized layers, such as improved switching speed and efficiency \cite{wang2013low,garello2014ultrafast}. The reduction of symmetry has been realized experimentally through a variety of means, such as the application of an in-plane magnetic field \cite{miron2011perpendicular,liu2012current,cubukcu2014spin}, the addition of in-plane magnetized layers (ferromagnetic \cite{lau2016spin,seung2018spin} or antiferromagnetic \cite{fukami2016magnetization,oh2016field}), or engineered structural asymmetry \cite{yu2014switching}.  Of relevance to this work is a series of recent experiments in which substrates with reduced crystal symmetry (transition metal dichalcogenide) were utilized to realize spin-orbit torques with a form that enables switching of perpendicular ferromagnets \cite{MacNeill2016,MacNeill2017,stiehl2019current,Stiehl2019}.
	
	While symmetry dictates the form of the spin-orbit torque, quantifying its magnitude and identifying its microscopic origin require explicit calculation.  Knowledge of these properties can aid in developing materials and device selection rules in order to optimize relevant figures of merit, such as switching efficiency.  With this motivation, we report on first-principles calculations of electric-field induced spin-orbit torque in transition metal dichalcogenide(TMD)-ferromagnet bilayers.  Our numerical analysis demonstrates that a considerable out-of-plane dampinglike torque is generated in this low symmetry system.  We also provide a general analysis of the symmetry-allowed torques for materials of this symmetry class.  We compute the spin current flowing between layers in the heterostructure, and find that it provides the primary source of torque on the magnetic layer.  
	
	Our paper is organized as follows. In Sec.~\ref{Methods} we review the procedure for computing spin-orbit torque using \textit{ab initio} methods and linear response theory. In Sec.~\ref{results} we present and discuss results obtained from the first-principles calculation of dampinglike and fieldlike spin-orbit torque. In Sec.~\ref{sec:spincurrent}, we compare the torque to the spin current flux, and in Sec. \ref{discussion} we present our conclusions.
	
	\section{Computational Methods}
	\label{Methods}
	\subsection{Formalism}
	The Hamiltonian of a magnetic system is generally given by the sum of a magnetization-dependent part and magnetization-independent part, which includes the kinetic energy $T$, potential energy $V$, and spin-orbit coupling.  Within the local spin density approximation (LSDA), the magnetic order leads to a spin-dependent exchange-correlation potential $\mathbf{\Delta}$ which couples to electron spin $\mathbf{S}$ through an exchange interaction $\mathbf{\Delta}\cdot\hat{\mathbf{S}}$. We approximate the spin-orbit coupling with an on-site, atomic-like form $\hat{\mathbf{L}}\cdot\hat{\mathbf{S}}$ \cite{fernandez2006site}.  The Hamiltonian is then given by:
	\begin{equation}
	\label{eq:Hamiltonian}
	H=T+V+\mathbf{\Delta}\cdot\hat{\mathbf{S}}+\alpha \hat{\mathbf{L}}\cdot\hat{\mathbf{S}},
	\end{equation}
	where $\hat{\mathbf{L}}$ and $\hat{\mathbf{S}}$ are (unitless) electron orbital angular momentum and spin Pauli matrices respectively, and $\alpha$ is a diagonal matrix parametrizing the spin-orbit coupling strength of different atoms.  The torque operator $\boldsymbol{\mathcal{T}}$ is calculated from the change of magnetization with respect to time,
	\begin{equation}
	\label{eq:torque}
	\boldsymbol{\mathcal{T}}=\frac{d\mathbf{\Delta}}{dt}=\frac{\mathrm{i}}{\hbar}[H,\mathbf{\Delta}]=-\frac{\mathrm{i}}{\hbar}[{\mathbf{\Delta}}\cdot\hat{\mathbf{S}} ,\hat{\mathbf{S}}].
	\end{equation}

	When an external electric field is applied, the electron distribution function and wave function are both modified. The linear response from the change of the Fermi-Dirac distribution function is time-reversal odd within the relaxation time approximation \cite{Haney2013}, while the linear response from the change of electronic wave function is time-reversal even \cite{Freimuth2014}. Using the Kubo formula \cite{Sinova2004,NagaosaRMP2010,Freimuth2014,SinovaRMP2015}, the even and odd torkances (the torque per electric field strength per area) under the external field in ${\bf x}$-direction are expressed as
	
	\begin{equation}
	\label{eq:torkance}
	\boldsymbol{\tau}^{\rm even}=2e \sum_{\substack{\mathbf{k},n\\
			m\neq n}} f(E_{n\bf{k}}) \frac{\text{Im}\bra{\psi_{n\bf{k}}}\frac{dH}{dk_x}\ket{\psi_{m\bf{k}}}\bra{\psi_{m\bf{k}}}\boldsymbol{\mathcal{T}}\ket{\psi_{n\bf{k}}}}{(E_{m\bf{k}}-E_{n\bf{k}})^2+\eta^2},
	\end{equation}
	\begin{equation}
	\label{eq:oddtorkance}
	\boldsymbol{\tau}^{\rm odd}=\frac{e}{\pi} \sum_{\substack{\mathbf{k},n,m}} \frac{\eta^2 \text{Re}[\bra{\psi_{m\bf{k}}}\frac{dH}{dk_x}\ket{\psi_{n\bf{k}}}\bra{\psi_{n\bf{k}}}\boldsymbol{\mathcal{T}}\ket{\psi_{m\bf{k}}}]}{[(\mu-E_{n\bf{k}})^2+\eta^2][(\mu-E_{m\bf{k}})^2+\eta^2]},
	\end{equation}
	where $\ket{\psi_{n\bf{k}}}$ satisfy the steady-state Schrodinger equation, $H_{\bf{k}}\ket{\psi_{n\bf{k}}}=E_{n\bf{k}}\ket{\psi_{n\bf{k}}}$, and are labeled by Bloch wave vector ${\bf k}$ and band index $n$.  In Eqs. \ref{eq:torkance}-\ref{eq:oddtorkance}, $e$ is the electron charge, $f(E)$ is the equilibrium Fermi-Dirac distribution function, $\mu$ is the chemical potential, and $\eta$ is the broadening parameter (note that $f$ depends on $\mu$). We use $k_BT=\eta=25~{\rm meV}$ throughout the paper unless otherwise noted.

	We assume $H$ is expressed in a tight-binding representation.  In evaluating Eqs.~\ref{eq:torkance} and ~\ref{eq:oddtorkance}, the velocity matrix operator is given by $\frac{dH}{dk_{x,y}}=\sum_{\bf{R}}(\mathrm{i}R_{x,y})e^{\mathrm{i}\bf{k}\cdot\bf{R}}H_{\bf{R}}$, where $H_{\bf{R}}$ is the Hamiltonian matrix coupling the orbitals centered in the primary unit cell with orbitals centered in the unit cell displaced by ${\bf R}$.  We include the atomic coordinate positions ({\it i.e.}, the basis vectors of atoms within the unit cell) in the displacement vector ${\bf R}$ \cite{Souza2019}.  Within the LSDA approximation, we can rotate the magnetization direction by rotating the time-reversal odd component of the real-space tight-binding Hamiltonian. 
	
	We also evaluate the spin current flowing between the two layers of the heterostructure, whose evaluation we describe next.  We write the Hamiltonian for the bilayer system as:
	\begin{eqnarray}
	H=\begin{pmatrix}
	H_{\rm FM}			&	t_{\rm FM,TMD} 	\\
	t_{\rm TMD,FM}	&	H_{\rm TMD}	\\
	\end{pmatrix}\label{eq:Hbilayer}
	\end{eqnarray}
	The diagonal elements of Eq. \ref{eq:Hbilayer} describe intralayer (or ``on-site'') contributions to the Hamiltonian.  The off-diagonal elements describe the coupling (or ``hopping'') between layers.  The interlayer electron current operator $J$ is given by:
	\begin{eqnarray}  
	J = i\begin{pmatrix}
	0			&  t_{\rm FM,TMD} 	\\
	-t_{\rm TMD,FM} 	& 0 	\\
	\end{pmatrix}.
	\label{eq:Qs}
	\end{eqnarray}
	We denote the spin current flowing between layers as ${\bf Q}$, whose vector direction specifies the spin polarization of the spin current. The spin current operator is the symmetrized product of the interlayer current and the spin operator:
	\begin{eqnarray}
	Q_\alpha = \frac{1}{2}\left( \hat{S}_\alpha J + J \hat{S}_\alpha\right) 
	\end{eqnarray}
	To evaluate the electric-field-induced spin current, we replace $\boldsymbol{\mathcal{T}}$ with ${\bf Q}$ in Eqs. \ref{eq:torkance} and \ref{eq:oddtorkance} \footnote{The code used to evaluate Eqs.~\ref{eq:torkance} and \ref{eq:oddtorkance} is available from the corresponding authors upon reasonable request.}.

	\subsection{First principles calculations}
	
	In this section we describe some details of the first principles calculations, which proceed in three steps: structural relaxation, Wannier projection, and evaluation of Eqs.~\ref{eq:torkance} and~\ref{eq:oddtorkance} using the tight-binding Hamiltonian.  
	
	The structural relaxation itself is a two-step process, in which we first relax using the Vienna Ab initio Simulation Package (VASP) \cite{VASP}\footnote{Disclaimer: Certain commercial products are identified in this paper in order to specify the theoretical procedure adequately. Such identification is not intended to imply recommendation or endorsement by the National Institute of Standards and Technology nor is it intended to imply that the software identified is necessarily the best available for the type of work.}, and use this relaxed structure as input to a second structural relaxation in Quantum ESPRESSO \cite{QE}. We find this is more efficient than exclusively using Quantum ESPRESSO for relaxation. In the VASP calculation, we use  the PAW method \cite{PAW} with the Perdew-Burke-Ernzerhof generalized-gradient approximation (PBE-GGA) functional \cite{PBEGGA}. All structures are relaxed until the total energy converges to within $10^{-4}~{\rm eV}$ during the self-consistent loop, with forces converged to $0.1~{\rm eV/nm}$, while employing the Methfessele-Paxton method with a smearing of $0.2~{\rm eV}$ width. The Brillouin zone is sampled with a $7 \times 4 \times 1$ Monkhorst Pack mesh \cite{MPmesh}. An energy cutoff of $450~{\rm eV}$ is used for all calculations. Van der Waals interactions are accounted for by means of the Grimme (D2) scheme \cite{Grimme}. The valence electron configurations for each metal considered are W: $5d^5 6s^1$, Te: $5s^2 5p^4$, Co: $3d^84s^1$ and Mo: $4d^5 5s^1$.

	\begin{figure}[t]
		\includegraphics[width=0.9\columnwidth]{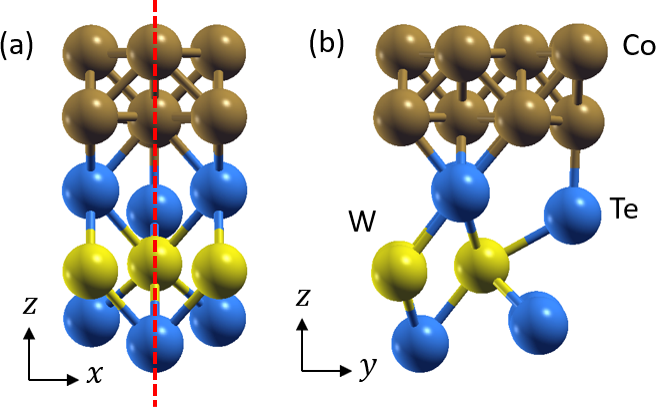}
		\caption{Depiction of system geometry for \ch{WTe2}(1)/Co(2).  Note the system retains mirror symmetry about the $yz$ plane $(x \rightarrow -x)$, but breaks mirror symmetry about the $xz$ plane. 
		}
		\label{Fig:geom2}
	\end{figure}
	
	Prior to creating the TMD/Co systems, the lattice constants of the isolated TMD layer are optimized, resulting in $a = 0.349~{\rm nm},~b=0.625~{\rm nm}$ for \ch{WTe2} and $a=0.348~{\rm nm},~b = 0.636~{\rm nm}$ for \ch{MoTe2}. The Co-TMD heterostructure consists of 1-2 monolayers of TMD stacked on a Co slab 3 layers thick with 4 Co atoms per layer.  There is a large mismatch between TMD and Co lattice constants.  In this work, we 
	apply most of the strain to the Co layer (see Fig. \ref{Fig:geom2}) since we focus on retaining the crystal symmetry of the TMD layer, and experimental samples exhibit a lack of crystallinity in the ferromagnetic layer \cite{MacNeill2016,MacNeill2017,Stiehl2019} with amorphous Permalloy as the ferromagnet layer.  More realistic treatments of the system would require a large TMD/Co unit cell to reduce the lattice mismatch. However such systems are not feasible due to the computational cost. To maintain the mirror symmetry ($x\rightarrow-x$) of the system a two-step optimization process is performed. In the first step all atomic positions are allowed to relax. In the following step the $x$ positions of the Co atoms are aligned with those of W (Mo) and Te and restricted from movement (frozen), and the structure is relaxed again.  The relaxed atomic positions are provided in Appendix~\ref{App.B}. 
	
	The relaxed structure provided by VASP serves as the initial configuration for the structural relaxation calculation in Quantum ESPRESSO.  The optimized Co atoms form three flat layers consisting of 4 atoms per layer. To reduce the tight-binding system size and computational load, we remove the top Co layer prior to relaxation in Quantum ESPRESSO. The relaxation calculation is nonrelativistic and spinpolarized, and utilizes a $12\times 6 \times 1 $ Monkhorst-Pack mesh \cite{MPmesh}, cutoff energy $1088~{\rm eV}$, total energy convergence threshold $1.36\times10^{-3}~{\rm eV}$, and force convergence threshold $2.57\times10^{-2}~{\rm eV/nm}$. 
	
	With the optimized geometry of the heterostructure and the corresponding self-consistent ground state computed with Quantum ESPRESSO, we use Wannier90 \cite{Wannier90} to obtain the real-space tight-binding model in the basis of atomic orbitals. We project onto $d$ orbitals of transition metal atoms, $s$ and $p$ orbitals of chalcogen atoms, and $s,p,d$ orbitals of Co atoms. After obtaining the collinear spin-polarized Hamiltonian in the Wannier basis, we add the onsite spin-orbit coupling terms $\alpha \hat{\mathbf{L}}\cdot\hat{\mathbf{S}}$. Note that adding spin-orbit coupling ``by hand'' in this manner requires that Wannier orbitals are not localized, in order to ensure that they are spherical harmonics consistent with the standard representation of $\mathbf{L}$. The strength of the spin-orbit coupling parameter $\alpha$ is obtained by fitting the pristine spin-orbit coupled bands from Quantum ESPRESSO.  We obtain $\alpha=\left[350,~100,~500,~70\right]~{\rm meV}$ for W, Mo, Te, and Co, respectively. We adopt this approach because it is technically easier to achieve a good Wannier projection of a collinear magnetized Hamiltonian, and the on-site spin-orbit coupling approximation yields accurate results. Note that the Wannier projection procedure breaks the mirror symmetry slightly, so we manually restore the mirror symmetry using procedures as described in Ref.~\cite{TBmodels}. We use a k mesh of approximately $540\times300$  (similar integration steps in the $k_x(k_y)$ direction) to evaluate Eqs.~\ref{eq:torkance} and \ref{eq:oddtorkance}.

	\section{Results}
	\label{results}
	
	\subsection{Symmetry considerations}
	
\begin{figure}[t]
		\includegraphics[width=1.\columnwidth]{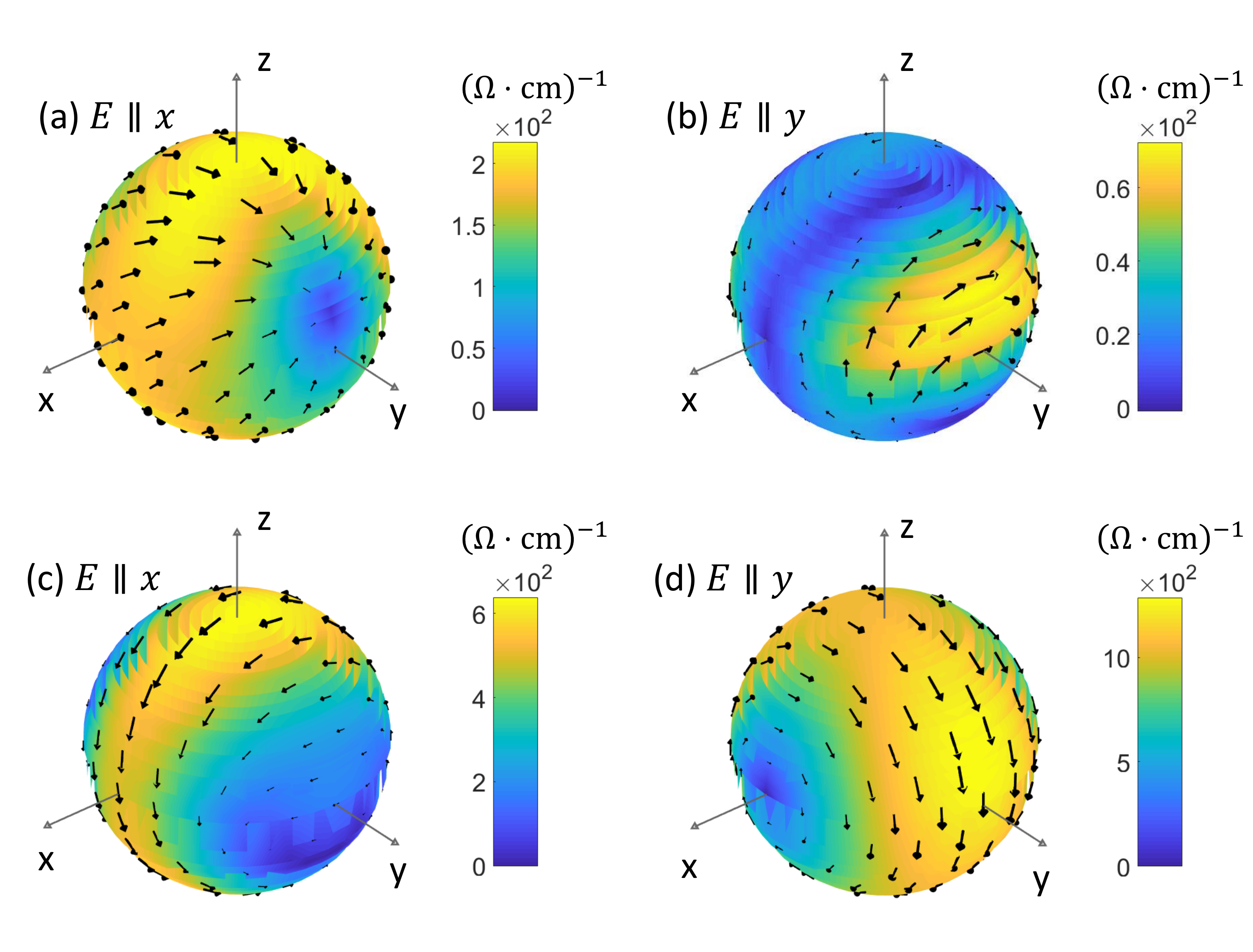}
		\caption{(Color online) Angular dependence of the dampinglike (a and b) and fieldlike (c and d) torkance on the magnetization direction $(\theta,\phi)$ for \ch{WTe2}(1)/Co(2) under an external electric field along the $\hat{x}$ (a and c) and $\hat{y}$ (b and d) direction at Fermi level. The heterostructure system keeps the mirror symmetry $x\rightarrow-x$ only. The arrow(color) on the sphere indicates the direction(magnitude) of the torkance under the given magnetization direction.
		}
		\label{Fig:GlobalTorque}
	\end{figure}

	We preface the presentation of numerical results with a review of the symmetry properties of the system, which have been discussed in previous works \cite{MacNeill2016,MacNeill2017,Stiehl2019}. In Appendix A, we present the fully general form of the current-induced torques for this system.  In what follows we focus on the azimuthal angle dependence for in-plane magnetization orientations (along the equator of Fig.~\ref{Fig:GlobalTorque}). We denote the out-of-plane and in-plane torques as follows:
	\begin{eqnarray}
	\mathcal{T}_{\perp}&=&\mathcal{T}_{z} \\
	\mathcal{T}_{\parallel}&=&\mathcal{T}_{x}\sin(\phi)-\mathcal{T}_{y}\cos(\phi),
	\end{eqnarray}
	where the azimuthal angle $\phi$ defines the in-plane magnetization direction: $M_x=M \cos{\phi}, M_y=M \sin{\phi}$. By examining the symmetry transformation of $\mathcal{T}_{\perp}$ and $\mathcal{T}_{\parallel}$ under an external field in the ${\bf x}$ direction, we find that both in-plane and out-of-plane torkances are cosine functions. We group them into time-reversal even and odd parts:
	\begin{eqnarray}
	\label{eq:fitting_Ex}
	\tau_{\perp}&=&\sum_{n=0}^\infty B_{2n}^{\perp,{\rm even}}\cos(2n\phi)+B_{2n+1}^{\perp,{\rm odd}}\cos\big((2n+1)\phi\big)\nonumber\\
	\tau_{\parallel}&=&\sum_{n=0}^\infty B_{2n}^{\parallel,{\rm odd}}\cos(2n\phi)+B_{2n+1}^{\parallel,{\rm even}}\cos\big((2n+1)\phi\big)\nonumber\\
	\end{eqnarray}
	In this work we use the terms ``time-reversal even torque'' and ``dampinglike torque'' interchangeably, and also use the terms ``time-reversal odd torque'' and ``fieldlike torque'' interchangeably.  The 1$^{\rm st}$ order contribution to the in-plane time-reversal even torque ($B^{\parallel,{\rm even}}_1$) is equivalent to the conventional dampinglike torque form, i.e., $\mathbf{M}\times \mathbf{M} \times ({\bf E}\times{\bf z})$, and the 0$^{\rm th}$ order contribution to the out-of-plane time-reversal even torque ($B^{\perp,{\rm even}}_0$) is the unconventional torque allowed only in the absence of mirror symmetry about the $xz$ plane.
	
	For an applied electric field in the ${\bf y}$-direction, the torques take the form
	\begin{eqnarray}
	\tau_{\perp}&=&\sum_{n}A_{2n}^{\perp,{\rm even}}\sin(2n\phi)+A_{2n+1}^{\perp,{\rm odd}}\sin\big((2n+1)\phi\big)\nonumber\\
	\tau_{\parallel}&=&\sum_{n}A_{2n}^{\parallel,{\rm odd}}\sin(2n\phi)+A_{2n+1}^{\parallel,{\rm even}}\sin\big((2n+1)\phi\big)\nonumber\\\label{eq:tau_Ey} 
	\end{eqnarray}
	For this direction of ${\bf E}$ field, mirror symmetry about the $yz$ plane is retained, and the torques assume a more conventional form.  In particular there is no magnetization-independent out-of-plane dampinglike torque.
	
	\subsection{Numerical Results}
	
	We next turn to the calculation results. Fig.~\ref{Fig:GlobalTorque} shows the torque as a function of magnetization orientation for a heterostructure composed of one \ch{WTe2} layer and two layers of Co (as shown in Fig.~\ref{Fig:geom2}). Figs.~\ref{Fig:GlobalTorque}(a) and (c) show the dampinglike and fieldlike components of torque, respectively, for ${\bf E}$ along ${\bf x}$. Recall that for this direction of ${\bf E}$, all symmetries are broken and the form of the torque is unconstrained. The dampinglike torque drives the magnetization into a point in the $yz$ plane.  As discussed in the introduction, for this case an applied electric field can deterministically switch a perpendicularly magnetized layer due to the dampinglike torque driving the magnetization to a point in the northern or southern hemisphere.  The fieldlike torque also vanishes at a point in the $yz$ plane, although at a point different than for which the dampinglike torque vanishes.  There is therefore no point at which the total electric-field-induced torque vanishes.
	
	Figs.~\ref{Fig:GlobalTorque}(b) and (d) show the same data for ${\bf E}$ along ${\bf y}$. This is a more conventional configuration in which there is a mirror plane symmetry with respect to the plane formed by ${\bf z}$ and ${\bf E}$. The dampinglike torque drives the magnetization into the ${\bf x}$ direction and vanishes there due to the mirror symmetry $x\rightarrow-x$.  However we note that the dampinglike torque has substantial contributions from higher-order terms, and is not well described by the lowest-order form $\left(\mathbf{M}\times \mathbf{M} \times {\bf x}\right)$ \cite{Belashchenko2019,sousa2020emergent}.  The fieldlike torque is well described by the simple form ${\bf M} \times {\bf x}$.

	\begin{figure}[htbp]
		\includegraphics[width=1\columnwidth]{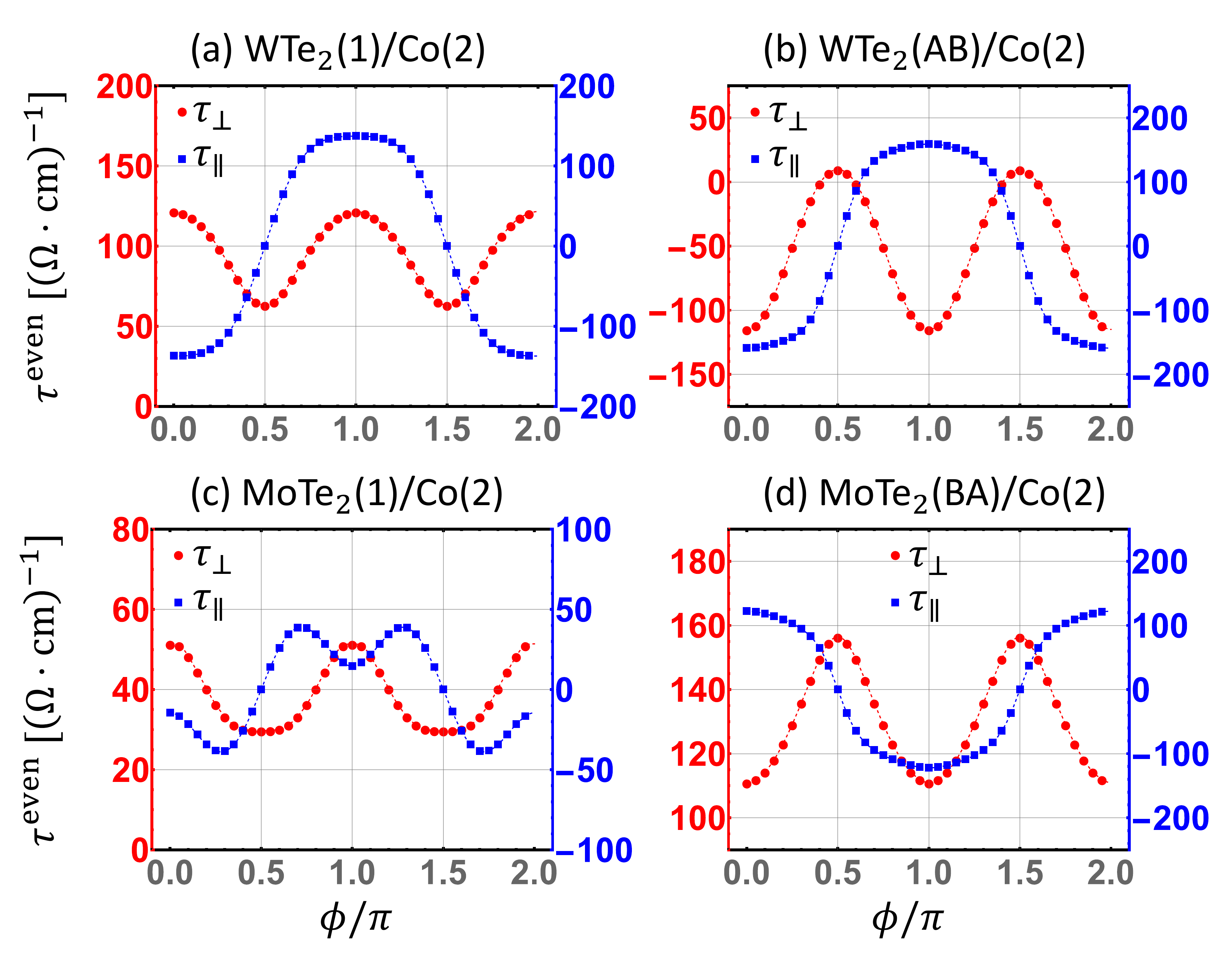}
		\caption{(Color online) Azimuthal angle($\phi$) dependence of torkances for (a)\ch{WTe2}(1)/Co(2), (b)\ch{WTe2}(AB)/Co(2), (c)\ch{MoTe2}(1)/Co(2), (d)\ch{MoTe2}(2)/Co(2) at Fermi level. Red circles and blue squares denote out-of-plane torkance $\tau_{\perp}\equiv\tau_z$ and in-plane torkance $\tau_{\parallel}\equiv\tau_{x}\sin(\phi)-\tau_{y}\cos(\phi)$, respectively. Dashed lines show the fitted results based on the symmetry-constrained form Eq.~\ref{eq:fitting_Ex} up to $n=9$.
		}
		\label{Fig:In-plane}
	\end{figure}
	
\begin{figure}[h]
		\includegraphics[width=.75\columnwidth]{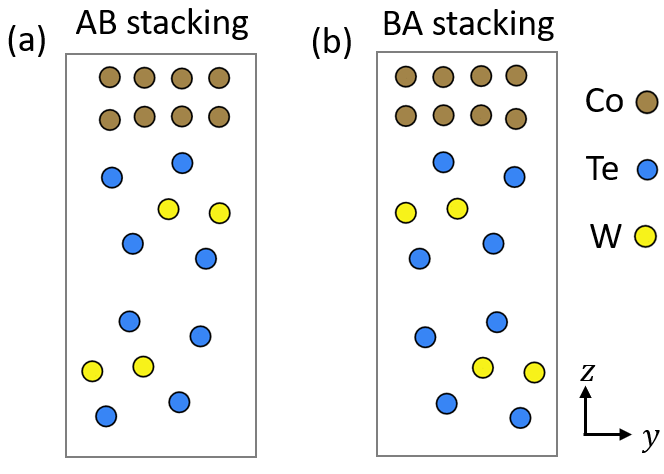}
		\caption{Panels (a) and(b) depict the atomic configurations for AB and BA stacking, respectively. }
		\label{Fig:ABBA}
	\end{figure}

	We next consider a series of systems in which we vary the TMD material type and thickness.  We restrict our attention to the torque as a function of azimuthal angle $\phi$ for in-plane magnetization directions, and only consider ${\bf E}$ along ${\bf x}$.  Fig.~\ref{Fig:In-plane} shows $\tau_{\parallel},~\tau_{\perp}$ versus $\phi$ for 1 and 2 layers of \ch{WTe2}, and 1 and 2 layers of \ch{MoTe2}, all with 2 layers of Co.  For the \ch{WTe2} structures, going from 1 to 2 layers changes the sign of the out-of-plane torque $\tau_\perp$ and decreases its magnitude. This is consistent with experimental observations \cite{MacNeill2017}, and indicates that, not surprisingly, the sign of $\tau_\perp$ is determined by the direction of in-plane symmetry breaking of the interfacial layer.  Consistent with this, we find that changing the order of stacking of \ch{WTe2} from AB to BA changes the sign of the out-of-plane torque (see Fig.~\ref{Fig:ABBA} for our definition of AB and BA stacking). AB and BA stacking cases are \textit{approximately} equivalent up to a mirror symmetry which flips the polarity of in-plane-symmetry-breaking order; however there are small differences resulting from the structural relaxation (the detailed atom locations of are provided in App.~\ref{App.B}). The reduction in magnitude for $\tau_\perp$ with additional layers of \ch{WTe2} can be expected because the two layers have opposite orientations of in-plane symmetry breaking. The out-of-plane torque due to the two layers should therefore exhibit partial cancellation, leading to a reduction in the overall magnitude.  We find that $\tau_{\parallel}$ is relatively insensitive to the \ch{WTe2} thickness.  
	
	For the \ch{MoTe2} structures, we find somewhat different behavior: $\tau_{\perp}$ and $\tau_{\parallel}$ both increase in magnitude going from 1 to 2 layers of \ch{MoTe2}.  It is surprising that $\tau_{\perp}$ increases with 2 layers, in light of the expected partial cancellation due to layers with opposite in-plane symmetry breaking, as described in the previous paragraph.  However, the torque is quite sensitive to other features of the structure, which we discuss later. Note that we do not calculate another AB stacking case of \ch{MoTe2} for simplicity since this will be similar to what we find in \ch{WTe2}.
	
	We fit our first-principles numerical results with the symmetry-constrained forms (Eq.~\ref{eq:fitting_Ex}) to extract the values of out-of-plane and in-plane torkances for five structures.  The two lowest-order torkance conductivities are summarized in Table.~\ref{Table:fitting}. Our numerical data conform to the symmetry-constrained forms with a sizable magnitude of constant out-of-plane torkance.  Note that the numerical data contain small but finite symmetry-allowed higher-order terms such as $\cos(2\phi)$ in the out-of-plane torkance.

	\begin{widetext}
		\begin{center}
			\begin{table}
				\begin{tabular}{|c|c|c|c|c|c|}					
					\hline					
					&\ch{WTe2}(1)/\ch{Co}(2) & \ch{WTe2}(AB)/\ch{Co}(2) & \ch{WTe2}(BA)/\ch{Co}(2) & \ch{MoTe2}(1)/\ch{Co}(2) & \ch{MoTe2}(BA)/\ch{Co}(2) \\\hline
					$B_0^{\perp,{\rm even}}$      &84$\pm$ 50  & -52   & 40   & 39    & 131 \\\hline
					$B_2^{\perp,{\rm even}}$      &30$\pm$ 28  & -63   & 65   & 11    & -22 \\\hline
					$B_1^{\parallel,{\rm even}}$  &-146$\pm$68 & -181  & -180 & -35   & 135 \\\hline
					$B_3^{\parallel,{\rm even}}$  &15$\pm$ 14  & 28    & 23   & 19    & -18 \\\hline
					$B_1^{\perp,{\rm odd}}$     &-301$\pm$410& 922   & 1165 & -1973 & -199\\\hline
					$B_3^{\perp,{\rm odd}}$     &-115$\pm$77 & 86    & -83  & -30   & -88 \\\hline 
					$B_0^{\parallel,{\rm odd}}$ &57$\pm$380  & -329  & 238  & 370   &  624\\\hline
					$B_2^{\parallel,{\rm odd}}$ &-107$\pm$81 & 74    & -79  & -56   & -29 \\\hline
					\hline
				\end{tabular}			
				\caption{Table of fitting parameters based on Eq.~\ref{eq:fitting_Ex} in five different heterostructures, given in units of $\hbar/2e~({\rm \Omega\cdot cm})^{-1}$. For the \ch{WTe2}(1) /\ch{Co}(2), we add an error bar from the Anderson disorder treatment described in the text. \ch{WTe2}(BA)/\ch{Co}(2)  and \ch{WTe2}(1)/\ch{Co}(2) share a similar interfacial geometry ({\it i.e.}, the same direction of in-plane symmetry breaking).
					\label{Table:fitting}}
			\end{table}
		\end{center}
	\end{widetext}
	
\begin{figure}[h]
		\includegraphics[width=1\columnwidth]{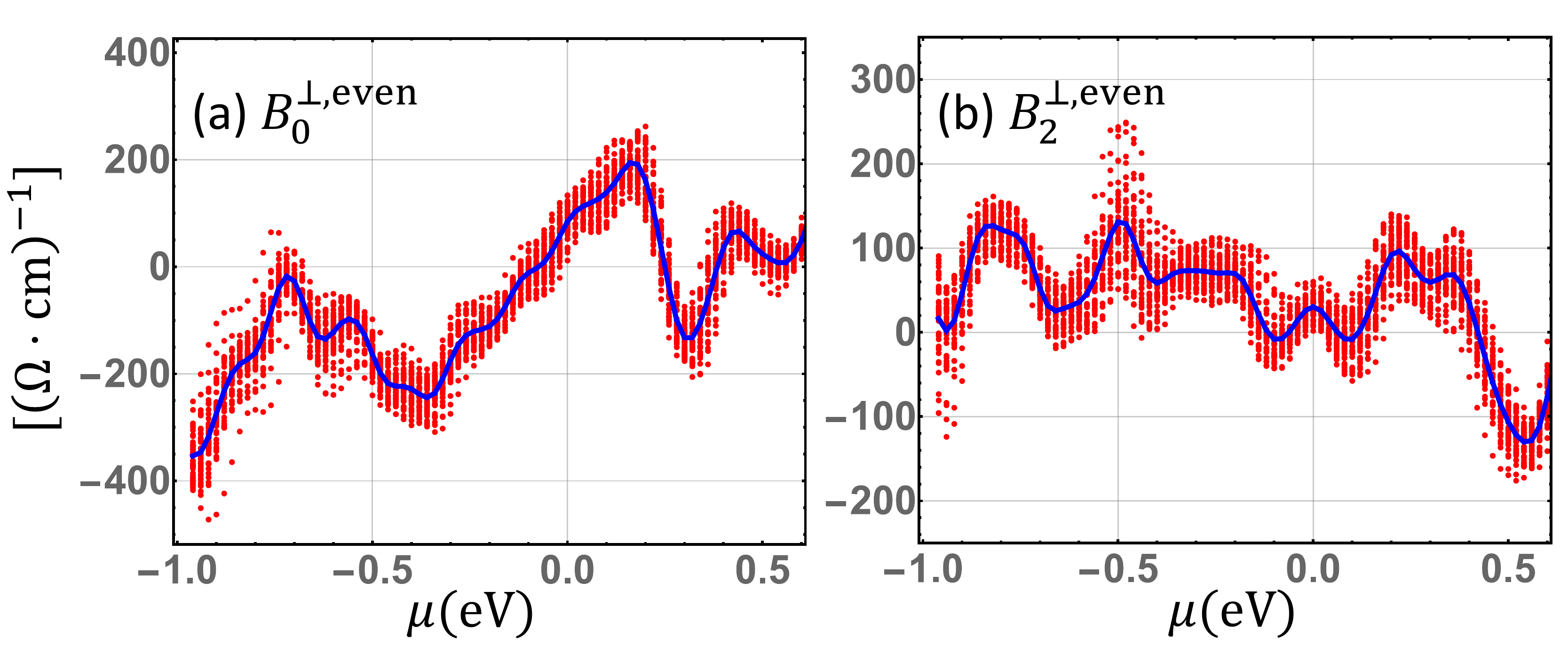}
		\caption{(Color online) Fitted parameters in \ch{WTe2}(1)/Co(2) as a function of chemical potential for 40 random onsite potentials $-0.1~\text{eV}<V_i<0.1~\text{eV}$ on all Co atoms. Panels (a) and (b) show lowest ($B^{\perp,{\rm even}}_0$) and second-lowest ($B^{\perp,{\rm even}}_2$) order terms for out-of-plane torque. Red dots denote all 40 possible outcomes at each chemical potential (relative to the Fermi energy) and blue lines show the average of all variations. }
		\label{Fig:Anderson}
	\end{figure}

	To test the robustness of the computed torkance values with respect to disorder, we utilize an Anderson disorder treatment \cite{Belashchenko2019} by adding a uniformly distributed random onsite potential $-V_m<V_i<V_m$ on all Co atoms. Fig.~\ref{Fig:Anderson} shows the torkance coefficients versus Fermi energy for 40 realizations of random on-site potentials for the \ch{WTe2}(1)/Co(2) system.  We find that adding a uniformly distributed random onsite potential with relatively small magnitude ($100$ meV) has a notable impact on the computed value.  The mean and standard deviation of the computed values are shown for WTe2(1)/Co(2) in Table \ref{Table:fitting}.  The standard deviation is substantial compared to the mean, underscoring the sensitivity of the computed values to details of the system. Note that we keep the chemical potential constant in different disorder realizations.  Increasing $V_m$ to $200$ meV results in similar values of torkance $( B_0^{\perp,\rm even}=94 \pm 64~\hbar/2e~({\rm \Omega\cdot cm})^{-1}, B_2^{\perp,\rm even}=32 \pm 56 ~\hbar/2e~({\rm \Omega\cdot cm})^{-1})$. Clearly, disorder plays an important role; however, a systematic study of the dependence of torkance on disorder is beyond the scope of the current paper.
	
	\section{Torque and Spin current} \label{sec:spincurrent}
	
	In heavy metal-ferromagnet bilayers, the dampinglike spin-orbit torque is conventionally viewed as a consequence of the spin current generated in the heavy metal through the spin Hall effect, which flows into the ferromagnet thereby exerting a torque on the magnetization.  First-principles calculations have shown that for Co-Pt bilayer systems, the dampinglike torque is indeed nearly equal to the spin current flux flowing between Pt and Co layers \cite{Freimuth2014,Mahfouzi2018,Belashchenko2019}.  Recent work has shown that there are other sources for spin-orbit torque \cite{Go2020}, such as the orbital Hall effect in the nonmagnet \cite{go2020orbital,OHE_TMD}, anomalous torque in the ferromagnet \cite{wang2019anomalous}, and interfacial torque \cite{manchon2009theory,amin2018interface}.  We next perform calculations to determine which mechanism applies for this set of systems.

\begin{figure}[h]
		\includegraphics[width=1\columnwidth]{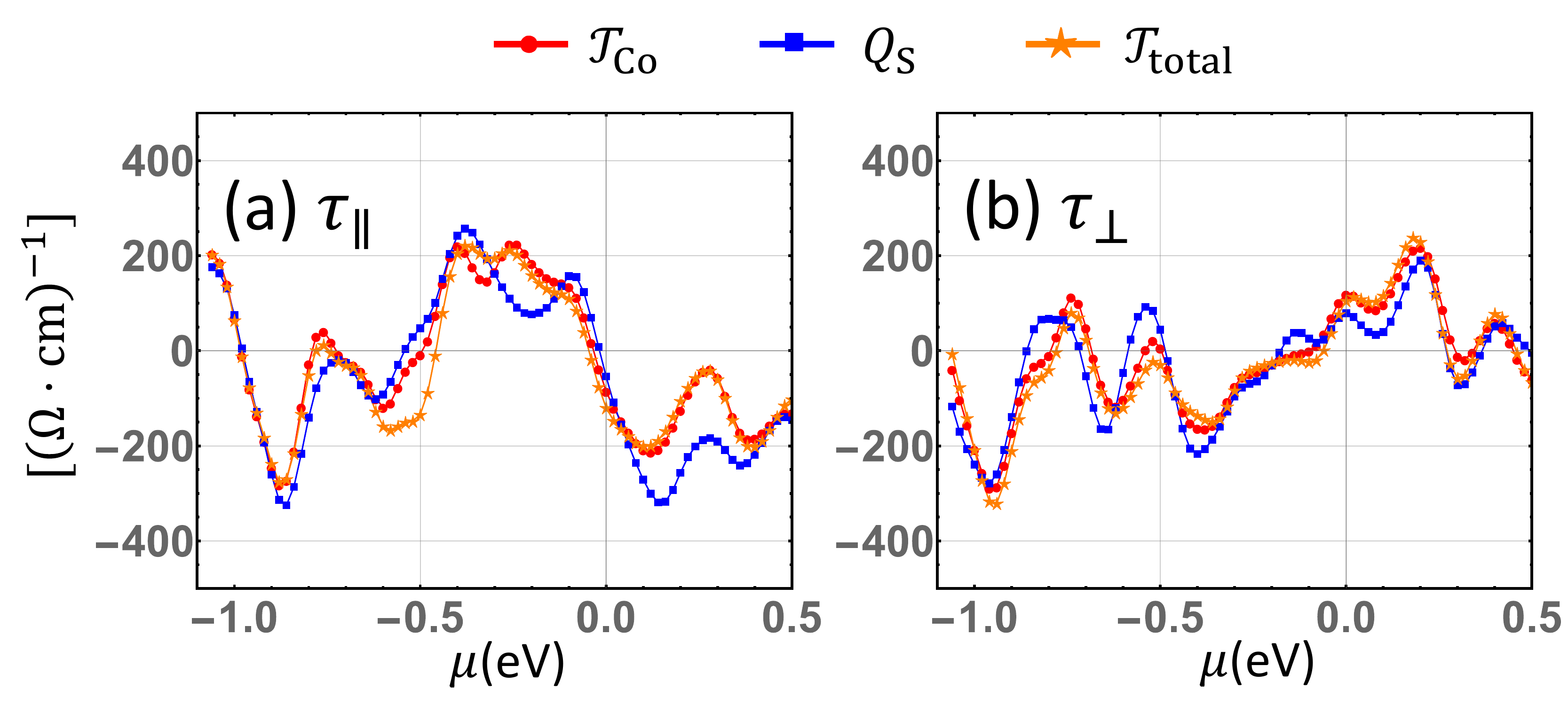}
		\caption{(a) In-plane torque $\tau_\parallel$ on Co-centered orbitals, total torque, and spin current with y polarization flowing between TMD and Co layers as a function of chemical potential.  (b) Same data for out-of-plane component of torque $\tau_\perp$ and spin current with z polarization. The electric field and magnetization are in the ${\bf x}$ direction.}
		\label{fig:Qs}
	\end{figure}

	Fig.~\ref{fig:Qs} shows that the torque on orbitals centered in the Co layer is approximately equal to the total torque, and, for most Fermi energies, also approximately equal to the spin current flux. There are regions of discrepancy between torque and spin current flux, particularly for the in-plane dampinglike torque $\tau_\parallel$ at higher Fermi energy.  We find that for this case, the difference between spin current and torque is due to spin-orbit coupling on the Co atoms, which acts as a drain on spin angular momentum and diverts incoming spin current into torque on the lattice \cite{haney2010current,Go2020}.  However generally there is good agreement between the spin current flux and the torque.  This indicates that the conventional picture of dampinglike torque arising from spin current flux generated in the substrate is generally applicable for this system.
	
	We note that the values obtained for the conventional dampinglike torque $B_1^{\parallel,{\rm even}}$ are similar in magnitude to the bulk spin Hall conductivity computed for \ch{WTe2} and \ch{MoTe2} \cite{Zhou2019}. 
	In the spin current + spin transfer torque picture, the unconventional out-of-plane dampinglike torque would arise from spin current flowing along the ${\bf z}$-direction (or $c$-axis of the TMD), with spin polarization along the ${\bf z}$-axis.  As shown in Ref \cite{MacNeill2016}, this component of the bulk spin Hall conductivity is symmetry-forbidden in crystals with the symmetry of \ch{WTe2}.  However, the non-symmorphic screw symmetry allows for a bulk spin current whose spin polarization along $\bf z$ alternates in sign between subsequent TMD layers ({\it i.e.}, a staggered spin current). This staggered response has been discussed in general terms in Ref. \cite{zhang2014hidden}, and realized in various contexts, such as in the staggered Rashba-Edelstein effect present in CuMnAs \cite{wadley2016electrical}.  We have performed preliminary calculations of this staggered out-of-plane spin Hall conductivity for bulk \ch{WTe2} and \ch{MoTe2} and compute conductivities which, although smaller than spin-orbit torque $B_0^{\perp,{\rm even}}$, are within the same order of a few tens of $(\hbar/2e)~({\rm \Omega\cdot  cm})^{-1}$. We leave this detailed analysis for future work.  The correspondence between torque and spin current indicates that maximizing the substrate material's bulk spin Hall conductivity is a viable strategy for maximizing the spin-orbit torque in the corresponding heterostructure. 
	
	The overall magnitude of the unconventional dampinglike torque we compute (see Table~\ref{Table:fitting}) is quite similar to the experimentally observed value of $\left(36\pm 8\right)$~$(\hbar/2e)~ ({\rm \Omega\cdot cm})^{-1}$ \cite{MacNeill2016}. This is substantially less than the value of dampinglike torque commonly observed and computed for the more conventional Co-Pt system, which is in the range $10^3~(\hbar/2e)~ ({\rm \Omega\cdot cm)^{-1}}$ \cite{liu2012current,Garello2013,Freimuth2014,nguyen2016spin,zhu2019spin}. In light of the correspondence between spin current and spin-orbit torque described above, this difference can be understood as a consequence of the relatively moderate magnitude of bulk spin Hall conductivity in \ch{WTe2} \cite{Zhou2019}.  We hypothesize that this is traced back to the large distance between successive TMD layers along the $c$ axis, but again leave a systematic analysis for future work.  We also note recent work which demonstrated magnetic switching using \ch{WTe2} substrates \cite{shi2019all}, indicating that the conventional dampinglike torque in these materials may be sufficiently strong for applications.  The magnitude of fieldlike torque we compute is also similar to that seen experimentally \cite{li2018spin,shao2016strong}. 
	
	\section{Conclusion}
	\label{discussion}
	
	In this work we presented first-principles calculations of the spin-orbit torque in a variety of TMD-Co bilayer systems.  As expected, the reduced symmetry of the TMD substrate enables novel forms of the spin-orbit torque, which can enable the deterministic switching of perpendicularly magnetized thin-film ferromagnets.  We find a magnitude of the dampinglike spin-orbit torque  which is consistent with experiment, and substantially less than that found in more commonly studied systems such as Co-Pt bilayers.  We also find that the spin-orbit torque is approximately equal to the spin current flux flowing from TMD layer to ferromagnetic layer, whose value is also commensurate with calculations of bulk spin Hall conductivity in TMD such as \ch{WTe2} and \ch{MoTe2}.  This suggests that maximizing the out-of-plane dampinglike torque may be accomplished by choosing substrate materials with large values of bulk out-of-plane spin-polarized spin Hall conductivity.

	\section{Acknowledgment}
	F.X. acknowledges support under the Cooperative Research Agreement between the University of Maryland and the National Institute of Standards and Technology Physical Measurement Laboratory, Award 70NANB14H209, through the University of Maryland.
	
	\appendix
	\section{General symmetry analysis}
	
	In this appendix we present the general form for the current-induced torque for a bilayer system whose only symmetry operation is mirror symmetry about the $yz$ plane.
	
	\subsection{Symmetry-constrained effective magnetic field}
	
	Assume that an applied electric field ($\bvec{E}$) gives rise to an effective magnetic field ($\bvec{B}$) in a ferromagnetic system.  To linear order in electric field, the response is given by 
	\begin{align}
	\bvec{B} = \chi(\mhat) \bvec{E} 
	\end{align}
	where $\chi$ is a $3 \times 3$ tensor that depends on the magnetization direction $\mhat = \big{(}\sqrt{1-z^2}\cos(\phi),\sqrt{1-z^2}\sin(\phi),z\big{)}$.  The torque on the magnetization is given by
	\begin{align}
	\bvec{\tau} = \bvec{\mhat} \times \bvec{B}
	\end{align}
	Since the unit vector $\mhat$ is parameterized by $z$ and $\phi$, the general response tensor can be written as
	\begin{align}
	\chi\TP =
	\begin{pmatrix}
	\chi_{11}\TP	&	\chi_{12}\TP		&	\chi_{13}\TP		\\
	\chi_{21}\TP	&	\chi_{22}\TP		&	\chi_{23}\TP		\\
	\chi_{31}\TP	&	\chi_{32}\TP		&	\chi_{33}\TP		\\
	\end{pmatrix}
	\end{align}
	Each component can be expanded in terms of real spherical harmonics,
	\begin{align}
	\chi_{ij}\TP &= \sum_{l=0}^{\infty} \sum_{m=0}^{l} P_l^m[z] \big{(} \alpha_{l,ij}^m \cos(m\phi) + \beta_{l,ij}^m \sin(m\phi) \big{)}
	\end{align}
	where $\alpha_{l,ij}^m$ and $\beta_{l,ij}^m$ are constant coefficients and $P_l^m$ are the associated Legendre polynomials. We rewrite this expression for convenience as
	\begin{align}
	\chi\TP = \sum_{l=0}^{\infty} \sum_{m=0}^{l} c^m_l\TP A^m_l + s^m_l\TP B^m_l
	\end{align}
	where
	\begin{align}
	c^m_l\TP &= P_l^m[z] \cos(m\phi) \\
	s^m_l\TP &= P_l^m[z] \sin(m\phi)
	\end{align}
	and
	\begin{align}
	A^m_l = 
	\begin{pmatrix}
	\alpha^m_{l,11}	&	\alpha^m_{l,12}	&	\alpha^m_{l,13}		\\
	\alpha^m_{l,21}	&	\alpha^m_{l,22}	&	\alpha^m_{l,23}		\\
	\alpha^m_{l,31}	&	\alpha^m_{l,32}	&	\alpha^m_{l,33}		\\
	\end{pmatrix},
	\quad
	B^m_l =
	\begin{pmatrix}
	\beta^m_{l,11}	&	\beta^m_{l,12}	&	\beta^m_{l,13}		\\
	\beta^m_{l,21}	&	\beta^m_{l,22}	&	\beta^m_{l,23}		\\
	\beta^m_{l,31}	&	\beta^m_{l,32}	&	\beta^m_{l,33}		\\
	\end{pmatrix}.
	\end{align}
	The purpose of rewriting the response tensor was to separate the functional dependence on magnetization direction ($z$,$\phi$) with the matrix structure ($A^m_{l}$,$B^m_{l}$).
	
	\subsection{Transforming the response tensor}
	
	In general, the response tensor transforms under some orthogonal transformation matrix $R$ as follows:
	\begin{align}
	\chi' = \det[R] R \chi R^\text{T}.
	\end{align}
	The $yz$ mirror plane transformation (i.e. $x \rightarrow -x$ for polar vectors) is given by
	\begin{align}
	R =
	\begin{pmatrix}
	-1	&	0	&	0	\\
	0	&	1	&	0	\\
	0	&	0	&	1	\\
	\end{pmatrix}
	\end{align}
	Under this transformation, $\chi$ becomes
	\begin{align}
	\chi' = -R \chi R^T =
	\begin{pmatrix}
	-\chi_{11}	&	\chi_{12}	&	\chi_{13}	\\
	\chi_{21}	&	-\chi_{22}	&	-\chi_{23}	\\
	\chi_{31}	&	-\chi_{32}	&	-\chi_{33}	\\
	\end{pmatrix}
	\end{align}
	To transform functions of magnetization, we note that the magnetization transforms like a pseudovector, which means that for a $yz$ mirror plane transformation, $z' = -z$ and $\phi' = -\phi$. Thus,
	\begin{align}
	c^m_l\TPp &= P_l^m[-z] \cos(-m\phi) = (-1)^{l+m} c^m_l\TP \\
	s^m_l\TPp &= P_l^m[-z] \sin(-m\phi) = -(-1)^{l+m} s^m_l\TP
	\end{align}
	where we have used the identity $P_l^m[-z] = (-1)^{l+m} P_l^m[z]$. To obtain the symmetrized response tensor, we simply add the transformed tensor to the original tensor, since $R$ and $I_{3\times3}$ are the only members of the symmetry group: 
	\begin{eqnarray}
	\chi_S\TP	&=& \chi\TP + \chi'\TPp \nonumber \\
	&=& \sum_{l=0}^{\infty} \sum_{m=0}^{l} 
	c^m_l\TP \Big{(} A^m_l + (-1)^{l+m+1} R A^m_l R^T \Big{)}
	+ \nonumber \\ && ~~~~~
	s^m_l\TP \Big{(} B^m_l + (-1)^{l+m} R B^m_l R^T \Big{)}
	\end{eqnarray}
	
	\subsection{Condensed form of response tensor}
	
	The response tensor derived in the last section contains matrices given by $A^m_l + (-1)^{l+m+1} R A^m_l R^T$ and $B^m_l + (-1)^{l+m} R B^m_l R^T$. Depending on the value of $l$ and $m$, these matrices take either of the following forms:
	\begin{align}
	S_l^m =
	\begin{pmatrix}
	a_l^m	&	0		&	0		\\
	0		&	e_l^m	&	f_l^m	\\
	0		&	h_l^m	&	i_l^m	\\
	\end{pmatrix}
	\quad
	T_l^m =
	\begin{pmatrix}
	0		&	b_l^m	&	c_l^m	\\
	d_l^m	&	0		&	0		\\
	g_l^m	&	0		&	0		\\
	\end{pmatrix}
	\end{align}
	Using this observation, we can rewrite the response tensor one last time in matrix form as
	\begin{align}
	\chi_S\TP = \sum_{l=0}^{\infty} \sum_{m=0}^{l} P_l^m[z] \big{(} s_l^m(\phi) S_l^m + t_l^m(\phi) T_l^m \big{)}
	\end{align}
	where
	\begin{align}
	s_l^m(\phi) =
	\begin{cases}
	\cos(m\phi)  &	\quad \text{for } l,m = \text{even/odd or odd/even}	\\
	\sin(m\phi)  &	\quad \text{for } l,m = \text{even/even or odd/odd}
	\end{cases}		\label{sDef} \\
	t_l^m(\phi) =
	\begin{cases}
	\sin(m\phi)  &	\quad \text{for } l,m = \text{even/odd or odd/even}	\\
	\cos(m\phi)  &	\quad \text{for } l,m = \text{even/even or odd/odd}
	\end{cases}. 	\label{tDef} 
	\end{align}
	We have arrived at the final form of the general response tensor. For a given $l$ and $m$, the contribution to each element of $\chi$ contains either $\sin(m\phi)$ or $\cos(m\phi)$ but not both.  This is the main consequence of the symmetry $x \rightarrow -x$.  Note that the matrix $T_0^0$ gives the magnetization-independent contribution ($l = 0, m = 0$) to $\chi$.  In general, the matrices $S_l^m$ and $T_l^m$ are odd and even with respect to the mirror plane transformation $x \rightarrow -x$ respectively.   
	
	\subsection{Symmetry-constrained torque}
	
	The torque can be written in terms of its own response tensor $\chi^T$ defined as follows 
	\begin{align}
	\bvec{\tau}	&= \bvec{\mhat} \times \bvec{B} \\
	&= \bvec{\mhat} \times \chi \bvec{E} \\
	&= \chi^T \bvec{E}
	\end{align}
	where $ \chi^T_{il} = \epsilon_{ijk} \hat{m}_j \chi_{kl}$.  Replacing the coordinate $z$ with $\theta$, where $z = \cos(\theta)$, one can then show that
	\begin{widetext}
		\begin{eqnarray}
		\chi^T(\theta,\phi) =	\sum_{l=0}^{\infty} \sum_{m=0}^{l} P_l^m[\cos(\theta)]
		\Big{(} 
		a_l^m A_l^m(\theta,\phi) + e_l^m E_l^m(\theta,\phi) + h_l^m H_l^m(\theta,\phi) +
		b_l^m B_l^m(\theta,\phi) + d_l^m D_l^m(\theta,\phi) + g_l^m G_l^m(\theta,\phi) 
		\Big{)} \nonumber \\\label{chiT}
		\end{eqnarray}
		where (assuming an in-plane electric field, i.e., $E_z = 0$)
		\begin{align}
		A_l^m &= s_l^m(\phi)
		\begin{pmatrix}
		0						&	0	&	0	\\
		\cos(\theta)			&	0	&	0	\\
		-\sin(\theta)\sin(\phi)	&	0	&	0	\\
		\end{pmatrix}
		\quad
		E_l^m = s_l^m(\phi)
		\begin{pmatrix}
		0	&	-\cos(\theta)			&	0	\\
		0	&	0						&	0	\\
		0	&	\sin(\theta)\cos(\phi)	&	0	\\
		\end{pmatrix}
		\quad
		H_l^m = s_l^m(\phi)
		\begin{pmatrix}
		0	&	\sin(\theta)\sin(\phi)	&	0	\\
		0	&	-\sin(\theta)\cos(\phi)	&	0	\\
		0	&	0						&	0	\\
		\end{pmatrix}
		\label{chiTMat1} \\
		B_l^m &= t_l^m(\phi)
		\begin{pmatrix}
		0	&	0						&	0	\\
		0	&	\cos(\theta)			&	0	\\
		0	&	-\sin(\theta)\sin(\phi)	&	0	\\
		\end{pmatrix}
		\quad
		D_l^m = t_l^m(\phi)
		\begin{pmatrix}
		-\cos(\theta)			&	0	&	0	\\
		0						&	0	&	0	\\
		\sin(\theta)\cos(\phi)	&	0	&	0	\\
		\end{pmatrix}
		\quad
		G_l^m = t_l^m(\phi)
		\begin{pmatrix}
		\sin(\theta)\sin(\phi)	&	0	&	0	\\
		-\sin(\theta)\cos(\phi)	&	0	&	0	\\
		0						&	0	&	0	\\
		\end{pmatrix}
		\label{chiTMat2} 
		\end{align}
	\end{widetext}
	Note that the assumption of an in-plane electric field means that the coefficients $c_l^m$, $f_l^m$, and $i_l^m$ are no longer relevant to the torque.  The expansion provided here captures all the consequences of symmetry, but is obviously quite complicated because there are only two symmetry operations in the symmetry group.  For each $l$ and $m$, there are six independent parameters given by $a_l^m$, $e_l^m$, $h_l^m$, $b_l^m$, $d_l^m$, and $g_l^m$.  The functions that contain the magnetization dependence are labeled accordingly.
	
	\subsection{Simpler expression for the out-of-plane torque}
	
	An out-of-plane ($z$) torque of some kind is required to switch ferromagnetic layers with perpendicular magnetic anisotropy. In the low-symmetry system we have studied here, such out-of-plane torques are nonvanishing. For an in-plane magnetization (i.e., $\theta = \pi/2$), the response tensor element relating an electric field along $\xhat$ with the out-of-plane torque $\tau_z$ is
	\begin{eqnarray}
	\chi^T_{zx}(\pi/2,\phi) &=&	\sum_{l=0}^{\infty} \sum_{m=0}^{l} P_l^m[0]
	\big{(} 
	a_l^m A_{l,zx}^m(\pi/2,\phi) +\nonumber\\ &&  ~~~~~~~~d_l^m D_{l,zx}^m(\pi/2,\phi)
	\big{)}	\\
	&=& \sum_{l=0}^{\infty} \sum_{m=0}^{l} P_l^m[0]
	\big{(} 
	- a_l^m s_l^m(\phi) \sin(\phi) + \nonumber \\ && ~~~~~~~d_l^m t_l^m(\phi) \cos(\phi)
	\big{)},
	\end{eqnarray}
	where we have used Eqs. \ref{chiT}, \ref{chiTMat1}, and \ref{chiTMat2}. We note that $P_l^m[0]$ is nonzero only when $l$ and $m$ are both even or both odd, which gives
	\begin{eqnarray}
	\chi^T_{zx}(\pi/2,\phi) &=& \sum_{l=0}^{\infty} \sum_{m}^{l}
	\big{(} 
	- a_{l}^m \sin(m\phi) \sin(\phi) + \nonumber \\&& ~~~~~~~d_{l}^m \cos(m\phi) \cos(\phi)
	\big{)} \\
	&=& \sum_{l=0}^{\infty} \sum_{m}^{l}
	\frac{1}{2} \big{(} 
	(d_l^m + a_l^m) \cos((m+1)\phi) + \nonumber \\&& ~~~~~~~(d_l^m - a_l^m) \cos((m-1)\phi)
	\big{)}
	\end{eqnarray}
	where $s_l^m(\phi)$ and $t_l^m(\phi)$ are replaced with $\sin(m\phi)$ and $\cos(m\phi)$ in the first line using Eqs. \ref{sDef} and \ref{tDef}. The last line is obtained using trigonometric summation identities. Note the second summation operator runs only over even (odd) values of $m$ for even (odd) values of $l$. 
	
	Finally, we note that the above expression contains several redundancies, because the $l$ dependence has dropped out and all integer multiples of $\phi$ are present somewhere in the sum. This yields
	\begin{align}
	\chi^T_{zx}(\pi/2,\phi) = \sum_{m=0}^{\infty} n_m \cos(m\phi) \label{tauZ}
	\end{align}
	where we have absorbed the redundant sums over coefficients $a_l^m$ and $d_l^m$ into $n_m$. 
	
	\section{Notes on First principles calculations details}
	\label{App.B}
	The atomic configurations of the relaxed structures are shown in the following tables. The unit of length for lattices vectors $a_i$ is the angstrom, and fractional coordinates are shown. 
	
	\begin{table}[htbp]
		\caption {Atomic Positions of \ch{WTe2}(1)/Co(2) and \ch{MoTe2}(1)/Co(2)}
		\begin{tabular}{cccc|cccc}
			\hline
			$a_1$&3.4895&0&0&~~$a_1$&3.4814&0&0\\
			$a_2$&0&6.254&0 &~~$a_2$&0&6.3562&0\\
			$a_3$&0&0&45    &~~$a_3$&0&0&39.9653 \\
			\hline
			W&0.0&0.0608~&0.2651~~~&~~Mo&0.0&0.3244~&0.3805\\
			W&0.5&0.4169~&0.2615~~~&~~Mo&0.5&0.6915~&0.3771\\
			Te&0.0&0.6639~&0.2954~~~&~~Te&0.5&0.4256~&0.4289\\
			Te&0.5&0.1541~&0.3088~~~&~~Te&0.0&0.9346~&0.4148\\
			Te&0.0&0.3212~&0.2170~~~&~~Te&0.5&0.0750~&0.3428\\
			Te&0.5&0.8085~&0.2311~~~&~~Te&0.0&0.5872~&0.3278\\
			Co&0.5&0.3240~&0.1720~~~&~~Co&0.5&1.0842~&0.2804\\
			Co&0.0&0.5821~&0.1720~~~&~~Co&0.5&0.5885~&0.2784\\
			Co&0.5&0.8215~&0.1755~~~&~~Co&0.0&0.3277~&0.2791\\
			Co&0.0&0.0642~&0.1727~~~&~~Co&0.0&0.8452~&0.2775\\
			Co&0.0&0.3218~&0.1354~~~&~~Co&0.5&0.3318~&0.2370\\
			Co&0.0&0.8239~&0.1340~~~&~~Co&0.0&0.5837~&0.2374\\
			Co&0.5&0.5827~&0.1344~~~&~~Co&0.5&0.8424~&0.2359\\
			Co&0.5&0.0643~&0.1351~~~&~~Co&0.0&1.0892~&0.2356\\
			\hline
		\end{tabular}
	\end{table}

	\begin{table*}[htbp]
		\caption {Atomic Positions of \ch{WTe2}(AB)/Co(2),\ch{WTe2}(BA)/Co(2), and \ch{MoTe2}(BA)/Co(2) }
		\begin{tabular}{cccc|ccc|cccc}
			\hline
			$a_1$&3.4895&0&0&3.4895&0&0&$a_1$&3.4814&0&0\\
			$a_2$&0&6.254&0&0&6.254&0&$a_2$&0&6.3562&0\\
			$a_3$&0&0&45&0&0&45&$a_3$&0&0&39.9653\\
			\hline
			W&0.0&0.0628~&0.4207~~~&0.5&0.9473~&0.4207~~~&~~Mo&0.5&0.1807~&0.5489\\
			W&0.5&0.9466~&0.2680~~~&0.0&0.5929~&0.4161~~~&~~Mo&0.0&0.8196~&0.5442\\
			W&0.0&0.5917~&0.2641~~~&0.0&0.0629~&0.2677~~~&~~Mo&0.0&0.3294~&0.3829\\
			W&0.5&0.4171~&0.4161~~~&0.5&0.4177~&0.2639~~~&~~Mo&0.5&0.6967~&0.3794\\
			Te&0.0&0.1992~&0.2337~~~&0.5&0.3436~&0.4505~~~&~~Te&0.0&0.0782~&0.5972\\
			Te&0.5&0.8134~&0.3869~~~&0.0&0.8508~&0.4642~~~&~~Te&0.5&0.5722~&0.5824\\
			Te&0.0&0.3211~&0.3725~~~&0.0&0.1968~&0.3869~~~&~~Te&0.0&0.4292~&0.5120\\
			Te&0.5&0.6877~&0.2196~~~&0.5&0.6895~&0.3725~~~&~~Te&0.5&0.9224~&0.4958\\
			Te&0.0&0.6667~&0.4506~~~&0.0&0.6655~&0.2973~~~&~~Te&0.5&0.4323~&0.4316\\
			Te&0.5&0.3441~&0.2975~~~&0.5&0.1579~&0.3117~~~&~~Te&0.0&0.9406~&0.4160\\
			Te&0.0&0.8515~&0.3120~~~&0.5&0.8104~&0.2335~~~&~~Te&0.5&0.0807~&0.3450\\
			Te&0.5&0.1589~&0.4642~~~&0.0&0.3218~&0.2194~~~&~~Te&0.0&0.5921~&0.3301\\
			Co&0.5&0.9424~&0.1750~~~&0.5&0.8206~&0.1779~~~&~~Co&0.5&1.0775~&0.2826\\
			Co&0.0&0.6844~&0.1746~~~&0.0&0.5807~&0.1744~~~&~~Co&0.5&0.5883~&0.2808\\
			Co&0.5&0.4258~&0.1749~~~&0.5&0.3225~&0.1744~~~&~~Co&0.0&0.3233~&0.2824\\
			Co&0.0&0.1848~&0.1780~~~&0.0&0.0632~&0.1751~~~&~~Co&0.0&0.8416~&0.2789\\
			Co&0.0&0.4250~&0.1372~~~&0.0&0.3204~&0.1378~~~&~~Co&0.5&0.3285~&0.2402\\
			Co&0.0&0.9428~&0.1374~~~&0.0&0.8227~&0.1364~~~&~~Co&0.0&0.5758~&0.2398\\
			Co&0.5&0.6843~&0.1380~~~&0.5&0.5814~&0.1369~~~&~~Co&0.5&0.8385~&0.2374\\
			Co&0.5&0.1845~&0.1367~~~&0.5&0.0630~&0.1375~~~&~~Co&0.0&1.0891~&0.2378\\
			\hline
		\end{tabular}
	\end{table*}

	\bibliography{reference}{}
	\bibliographystyle{apsrev4-1}

\end{document}